\documentclass[superscriptaddress,aps,9pt,twocolumn]{revtex4-2}
\usepackage[utf8]{inputenc}
\usepackage{graphicx}
\usepackage{verbatim}
\usepackage{amsmath}
\usepackage{amssymb}
\usepackage{bm}
\usepackage{color}
\usepackage{braket}
\usepackage[linesnumbered,ruled,vlined]{algorithm2e}
\SetKwInput{KwInput}{Input}                
\SetKwInput{KwOutput}{Output}              

\SetCommentSty{mycommfont}
\begin{document}
\title{Computing Free Energies with Fluctuation Relations on Quantum Computers}
\author{Lindsay Bassman}
\affiliation{Lawrence Berkeley National Lab, Berkeley, CA 94720}
\author{Katherine Klymko}
\affiliation{Lawrence Berkeley National Lab, Berkeley, CA 94720}
\author{Diyi Liu}
\affiliation{Department of Mathematics, University of Minnesota, MN 55455}
\author{Norm M. Tubman}
\affiliation{NASA Ames Research Center, Mountain View, CA 94035}
\author{Wibe A. de Jong}
\affiliation{Lawrence Berkeley National Lab, Berkeley, CA 94720}

\begin{abstract}
Fluctuation relations allow for the computation of equilibrium properties, like free energy, from an ensemble of non-equilibrium dynamics simulations. Computing them for quantum systems, however, can be difficult, as performing dynamic simulations of such systems is exponentially hard on classical computers.  Quantum computers can alleviate this hurdle, as they can efficiently simulate quantum systems.  Here, we present an algorithm utilizing a fluctuation relation known as the Jarzynski equality to approximate free energy differences of quantum systems on a quantum computer.  We discuss under which conditions our approximation becomes exact, and under which conditions it serves as a strict upper bound.  Furthermore, we successfully demonstrate a proof-of-concept of our algorithm using the transverse field Ising model on a real quantum processor.  The free energy is a central thermodynamic property that allows one to compute virtually any equilibrium property of a physical system.  Thus, as quantum hardware continues to improve, our algorithm may serve as a valuable tool in a wide range of applications including the construction of phase diagrams, prediction of transport properties and reaction constants, and computer-aided drug design in the future.  
\end{abstract}

\maketitle

\textit{Introduction.--}Thermodynamics is one of the most well-established and powerful physical theories, with impacts ranging from deep concepts, such as the arrow of time, to practical and technological applications, like the steam engine.  Its ability to compute bulk properties of macroscopic systems stems from dealing with averages over very large numbers of particles where individual deviations from the mean become insignificant.  
However, as system sizes decrease down to microscopic scales, these deviations, or fluctuations, from the average become appreciable.  
In finite temperature systems, thermal motion is the main source of fluctuations, while in zero- and low-temperature systems quantum effects begin to play an important role.  Regardless of their source, when fluctuations about the average become significant, classical thermodynamics begins to lose accuracy and it becomes necessary to apply stochastic thermodynamics. 

Stochastic thermodynamics allows thermodynamic concepts such as work and heat to be defined in terms of the statistics of an ensemble of trajectories of the system \cite{crooks1999excursions, seifert2012stochastic, strasberg2019operational}.  This framework has led to the discovery of fluctuation relations \cite{campisi2011colloquium, spinney2012fluctuation}, which relate fluctuations in non-equilibrium processes to equilibrium properties like the free energy.  
Arguably the most celebrated fluctuation relation is the Jarzynski equality \cite{jarzynski1997equilibrium, jarzynski1997nonequilibrium}, in which the free energy difference between two equilibrium states of a system is derived from an exponential average over an ensemble of measurements of the work required to drive the system from one state to the other.  While the Jarzynski equality was initially derived and experimentally verified for classical systems \cite{hummer2001free, liphardt2002equilibrium, douarche2005experimental, harris2007experimental, saira2012test, toyabe2010experimental}, it has since been extended to the quantum regime, for both closed~\cite{tasaki2000jarzynski, kurchan2000quantum, piechocinska2000information, mukamel2003quantum, monnai2005unified, esposito2009nonequilibrium, campisi2011colloquium} and open systems \cite{yukawa2000quantum, chernyak2004effect, de2004quantum, crooks2008jarzynski, campisi2009fluctuation, ngo2012demonstration, cuetara2015stochastic, venkatesh2015quantum, sone2020jarzynski}.  Experimental verification of the quantum Jarzynski equality was proposed \cite{huber2008employing} and later demonstrated with a liquid-state nuclear magnetic resonance platform \cite{batalhao2014experimental} and with cold trapped-ions \cite{an2015experimental}.

While the Jarzynski equality has proven important theoretically, providing one of the few strong statements that can be made about general non-equilibrium systems, its utility for computing free energies of relevant quantum systems has thus far been limited.  This is because simulating the exact trajectories of quantum systems on classical computers requires resources that scale exponentially with system size.  Therefore, computing even a single trajectory of a quantum system with tens of particles can quickly become intractable on classical computers, let alone an ensemble of trajectories.  

One potential path forward for computing this ensemble of trajectories is to employ quantum computers, which can efficiently simulate the dynamics of quantum systems \cite{feynman1982simulating, lloyd1996universal, abrams1997simulations, zalka1998simulating}.  Above and throughout the rest of the paper we use the term trajectory to evoke the correspondence with classical stochastic thermodynamics but we note that the classical notion of a trajectory is not directly applicable to quantum systems.  We instead use the term trajectory as a mathematical tool to define physical quantities that are given by averages over single realizations of physical processes~\cite{deffner2013quantum}.

Here, we present an algorithm to approximate free energy differences using the Jarzynski equality on a quantum computer.  We discuss under which conditions the approximation becomes exact and under which conditions the approximation provides a strict upper bound, which is tighter than the usual upper bound provided by the reversible work theorem.  We provide a proof-of-concept for our algorithm by computing free energy differences of a transverse field Ising model (TFIM) on a real quantum processor. By capitalizing on the ability of quantum computers to efficiently perform dynamic simulations, our algorithm enables the approximation of thermodynamic properties of quantum systems, allowing for deeper exploration into the evolving field of quantum thermodynamics.

\begin{figure}[ht]
	\centering
	\includegraphics[scale=0.53]{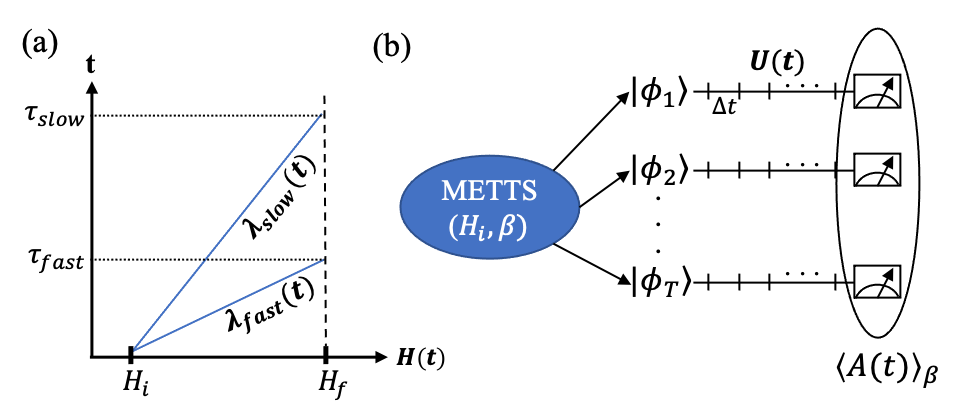}
	\caption{(a) Schematic diagram depicting how the parameter-dependent Hamiltonian $H(\lambda)$ can be varied over different total times $\tau$. (b) Schematic diagram for the METTS protocol.  The protocol requires as input the Hamiltonian and the inverse temperature $\beta$ of the equilibrium system.  The protocol generates a Markov chain of pseudo-thermal states $\phi_i$, which can be time-evolved under a separate Hamiltonian, measured, and averaged over to produce the thermal average for some time-dependent observable $A(t)$ at inverse temperature $\beta$.} 
	\label{fig:schematics}
\end{figure}

\textit{Theoretical Background and Framework.--}To use the Jarzynski equality in practice, we define a parameter-dependent Hamiltonian for the system of interest $H(\lambda)$, where $\lambda$ is an externally controlled parameter that can be adjusted according to a fixed protocol.  The Jarzynski equality uses work measurements from an ensemble of trajectories as $\lambda$ is varied to compute the free energy difference between the initial and final equilibrium states.  The equality is given by
\begin{equation}
\label{JE}
    e^{-\beta\Delta F} = \langle e^{-\beta W} \rangle , 
\end{equation}
where $\beta = \frac{1}{k_\text{B} T}$ is the inverse temperature $T$ of the system ($k_\text{B}$ is Boltzmann's constant) in its initial equilibrium state, $\Delta F$ is the free energy difference between the initial and final equilibrium states, $W$ is the work measured for a single trajectory, and $\langle...\rangle$ represents taking an average over the ensemble of trajectories. Without loss of generality, we assume the initial Hamiltonian of the system $H_i = H(\lambda = 0)$, and the final Hamiltonian $H_f = H(\lambda = 1)$.  As shown in Figure \ref{fig:schematics}a, the protocol for varying $\lambda$, denoted $\lambda(t)$, takes place over a time $\tau$, which can be defined to be as fast or slow as desired.  In general, the faster the protocol, the more trajectories will be required to compute a more accurate estimate of the free energy \cite{oberhofer2005biased} (see Section I in the Supplemental Information (SI)).

The main challenge in implementing such a procedure is preparing the initial thermal state on the quantum computer.  This is a non-trivial problem for which only a handful of algorithms have been proposed, most of which are either not NISQ-friendly (in terms of circuit complexity) or struggle to scale to large or complex systems \cite{poulin2009sampling, riera2012thermalization, verdon2019quantum, wu2019variational, zhu2020generation}. A method that is particularly promising for NISQ computers produces a Markov chain of sampled pseudo-thermal states, known as minimally entangled typical thermal states (METTS) \cite{white2009minimally, stoudenmire2010minimally}.  Averages of observables over the ensemble of METTS will converge to the true thermal average of the observable with increasing sample size.  The quantum version of the METTS protocol, known as QMETTS, gives a procedure to generate METTS on quantum computers using quantum imaginary time evolution (QITE) \cite{motta2020determining}. While initially presented as an algorithm to obtain thermal averages of static observables, METTS has been shown to enable calculation of time-dependent quantities~\cite{bonnes2014light}, and recently QMETTS was used to measure dynamic observables on quantum hardware~\cite{sun2021quantum}.  This can be achieved by evolving the METTS in real-time before measurement.  Figure \ref{fig:schematics}b shows schematically how measurements of a time-dependent observable $A(t)$ can be averaged over an ensemble of time-evolved METTS for a system with Hamiltonian $H_i$ at inverse temperature $\beta$ to give the thermal expectation value of $A(t)$. See Section II in the SI for more details.

Given an ensemble of pseudo-thermal states generated with the METTS protocol, the initial thermal energy $\langle E_i \rangle$ of the system at inverse temperature $\beta$ can be measured by averaging over energy measurements of the individual states in the ensemble.  Similarly, the final thermal energy $\langle E_f \rangle$ of the system after the $\lambda$ protocol has been implemented can be computed by time-evolving each pseudo-thermal state in the ensemble under the $\lambda$ protocol and averaging over energy measurements of the individual time-evolved states.  Now, for closed quantum systems, the work performed in a process is given by the difference in energy of the system before and after the process; therefore, the average thermal energies computed with the METTS ensemble can be used to compute the average work performed over the $\lambda$ protocol, as $\langle W \rangle$ = $\langle E_f \rangle$ - $\langle E_i \rangle$.  Note that $\langle W \rangle$ is always an upper bound on the free energy difference due to the reversible work theorem.  However, we endeavor to obtain a better approximation to the free energy difference by considering the distribution of individual pseudo-work values derived from the METTS ensemble.  

In this framework, we let each METTS in the ensemble correspond to a trajectory. For each trajectory, we compute a pseudo-work value by taking the difference of the measured initial and final energies of the sampled pseudo-thermal state before and after evolving it under the $\lambda$ protocol, respectively.  While the average over this ensemble of pseudo-work values will converge to the correct value for average work, the individual values in the ensemble are not necessarily physical work values. This is because the METTS protocol only guarantees accurate averages over the ensemble of METTS.  Nevertheless, we show that this distribution can be used in the Jarzynski equality to compute an approximate free energy difference $\Delta\Tilde{F}$ as
\begin{equation}
\label{FE_approx}
    e^{-\beta \Delta\widetilde{F}} = \langle e^{-\beta \widetilde{W}} \rangle , 
\end{equation}
where $\widetilde{W}$ are the individual pseudo-work values computed with the METTS ensemble.  In the limit of $\beta \rightarrow 0$, this approximation to the free energy difference becomes exact.  In the limit of $\beta \rightarrow \infty$, $\Delta\widetilde{F}$ is exact for $\lambda$ protocols that are adiabatic.  For arbitrary $\beta$, $\Delta\widetilde{F}$ upper bounds the true $\Delta F$ for adiabatic $\lambda$ protocols, and is a better approximation to the free energy difference than $\langle W \rangle$ due to Jensen's inequality.  See Sections III and IV of the SI for proofs of these statements.  For non-adiabatic $\lambda$ protocols, we empirically find that $\Delta F \le \Delta\widetilde{F} \le \langle W \rangle$ for a range of $\beta$'s and spin-model Hamiltonians, see Section V of SI.  Plugging the pseudo-work distribution into the Jarzynski equality, therefore provides a very good approximation to the free energy difference for closed quantum systems under certain conditions, and provides a tighter upper bound to the free energy difference than the average work in a broad range of cases.  We emphasize that while our algorithm can only approximate the free energy difference, it is one of the very few algorithms that can feasibly be performed on near-term quantum computers \cite{francis2020many}; and in many instances, this approximation can provide a strict, and even tight, upper bound on the free energy difference.

\textit{Algorithm.--}
We now describe our algorithm, which provides a procedure for obtaining a pseudo-work distribution from non-equilibrium dynamic simulations of a closed quantum system on a quantum computer, which can be used to approximate free energy differences.   
\begin{algorithm}
\DontPrintSemicolon
\label{algo:pseudocode}
  \KwInput{$H(\lambda$), $\beta$, $\lambda(t)$, $M$}
  \KwOutput{Free energy difference}
  work$\_$distribution = [ ]\\
  METTS$\_$state = random$\_$product$\_$state()\\
  \tcc{Loop over M trajectories}
  \For{m=[0,M]}
  {
      \tcc{make thermal state preparation circuit}
        circ$\_$TS = TS$\_$circ($\beta$, $H(\lambda=0)$, METTS$\_$state)\\
        \tcc{get initial state for next trajectory}
        METTS$\_$state = collapse(circ$\_$TS, m)\\
        \tcc{measure inital energy}
        $E_i$ = measure(circ$\_$TS, $H(\lambda=0)$) \\
        \tcc{make Hamiltonian evolution circuit}
        circ$\_$hamEvol = hamEvol$\_$circuit($\lambda(t)$, $H(\lambda)$)\\
        \tcc{measure final energy}
        total$\_$circ = circ$\_$TS + circ$\_$hamEvol \\
        $E_f$ = measure(total$\_$circ, $H(\lambda=1)$) \\
        work = $E_f$ - $E_i$\\
        work$\_$distribution.append(work)\\
  }
  \textbf{return} compute$\_$free$\_$energy(work$\_$distribution, $\beta$)
\caption{Pseudocode for approximation of free energy differences using METTS with the Jarzynski equality on quantum computers.}
\end{algorithm}
Pseudocode is shown in Algorithm \ref{algo:pseudocode}.  The algorithm takes as input the parameter-dependent Hamiltonian $H(\lambda)$, the inverse temperature $\beta$ of the initial system at equilibrium, the protocol $\lambda(t)$ to evolve the parameter from 0 to 1, and the total number of trajectories $M$.  The algorithm generates a pseudo-work distribution by looping over the $M$ trajectories.  For each trajectory, a circuit is generated which prepares the sampled pseudo-thermal state at inverse temperature $\beta$ ($circ\_TS$).  According to the QMETTS protocol, this is accomplished by initializing the qubits into an initial product state ($METTS\_state$) and which is then evolved under the initial Hamiltonian for an imaginary time $\beta/2$ using QITE.  For the first trajectory, $METTS\_state$ is a random product state, while for all subsequent trajectories $METTS\_state$ is determined by a projective measurement of $circ\_TS$ from the previous trajectory. This projective measurement happens next, by collapsing $circ\_TS$ into a basis which depends on the parity of trajectory $m$.  In order to ensure ergodicity and reduce auto-correlation times, it is helpful to switch between measurement bases throughout sampling \cite{stoudenmire2010minimally}. Following the method proposed in Ref.~\cite{stoudenmire2010minimally} for spin-$\frac{1}{2}$ systems, we measure (i.e., collapse) along the $z-$axis for odd trajectories, while for even trajectories we measure along the $x-$axis.

The thermal state circuit $circ\_TS$ is then used twice more to measure initial and final energies.  The initial energy $E_i$ is obtained by measuring the expectation value of the initial Hamiltonian $H_i$ in the pseudo-thermal state.  Next, a circuit ($circ\_hamEvol$) is generated to evolve the system under the time-dependent Hamiltonian according to the $\lambda$ protocol.  $circ\_TS$ and $circ\_hamEvol$ are concatenated to make $total\_circ$, which generates the time-evolved pseudo-thermal state.  The final energy $E_f$ is then obtained by measuring the expectation value of the final Hamiltonian $H_f$ in this state.  The difference of the two energies gives the pseudo-work for the given trajectory.  The free energy difference can then be approximated by plugging the pseudo-work ensemble into Eq.~\eqref{FE_approx}.
\begin{figure}[ht]
	\centering
	\includegraphics[scale=0.65]{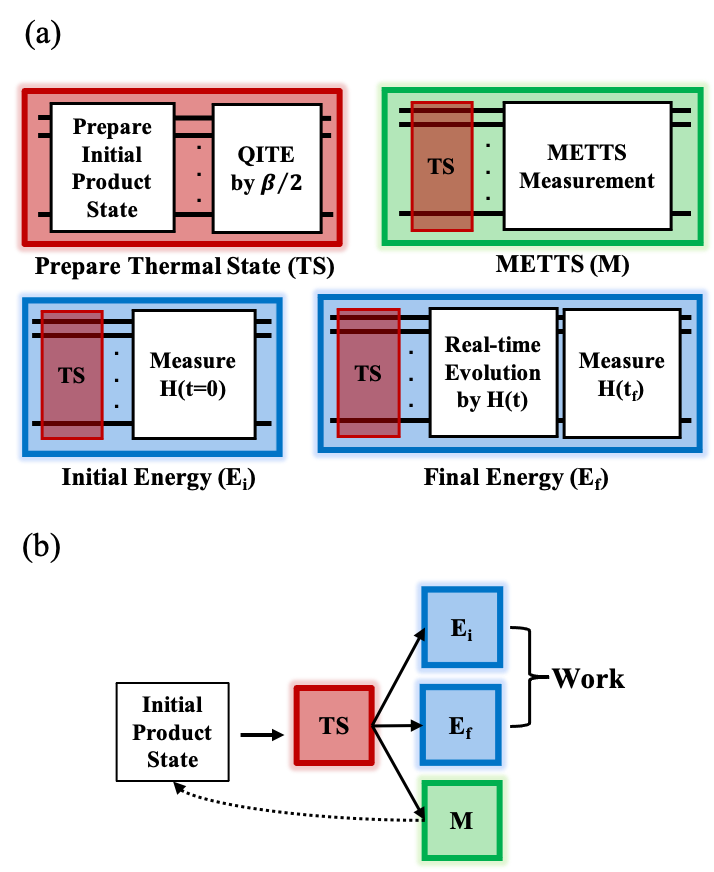}
	\caption{Circuits generated and workflow diagram for the algorithm.  (a) Quantum circuit diagrams for the thermal state preparation circuit (red), which is used in three separate circuits for measuring the initial and final energies (blue) as well as a circuit to measure the initial product state for the subsequent trajectory (green).  (b)  Workflow diagram depicting how the circuits above are integrated to produce a pseudo-work distribution.}
	\label{fig:workflow}
\end{figure}

Figure \ref{fig:workflow} depicts the quantum circuits that must be generated and the workflow diagram for how they must be executed.  The thermal state (TS) preparation circuit (i.e., $circ\_TS$) is depicted in Figure \ref{fig:workflow}a.  The circuit is embedded into three separate circuits: one to generate the next sampled state according to the METTS procedure ($M$), and two more for computing a pseudo-work value. The initial energy measurement circuit ($E_i$) simply composes the TS circuit with a set of gates that measures the value of the initial Hamiltonian, $H(\lambda=0)$.  The final energy measurement circuit ($E_f$) composes the $TS$ circuit with real-time evolution of the system as $H(\lambda)$ is varied according to the $\lambda$ protocol, as well as a set of gates that measures the value of the final Hamiltonian, $H(\lambda=1)$.  Figure \ref{fig:workflow}b shows how a pseudo-work value is derived from these circuits for each trajectory and how measurement of the $M$ circuit from the previous trajectory provides input to the $TS$ circuit for the next trajectory.  Note that the first few trajectories should be discarded as ``warm-up" values~\cite{white2009minimally}.

\textit{Results.--}
\begin{figure}[ht]
	\centering
	\includegraphics[scale=0.7]{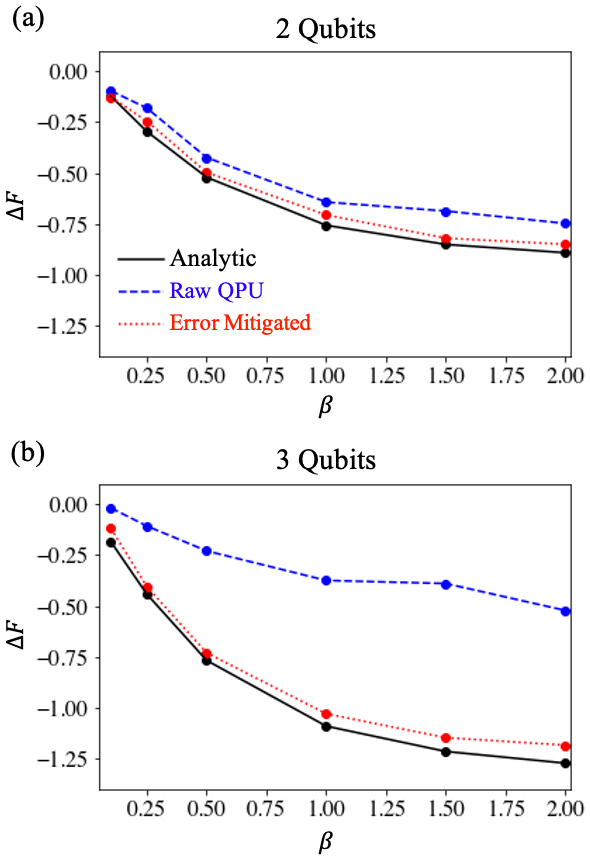}
	\caption{Approximate free energy differences ($\Delta \widetilde{F}$) for 2- and 3-qubit systems initialized at various inverse temperatures $\beta$ performed on an IBM quantum processing unit (QPU). The solid black line give the analytically computed values ($\Delta{F}$) for reference.  The blue dashed lines show raw results from the QPU, while the red dotted lines show these results after error mitigation has been performed.}
	\label{fig:hardware_results}
\end{figure}
We demonstrate our algorithm on real quantum hardware with a 2- and 3-qubit TFIM as a proof-of-concept.  The Hamiltonian is defined as
\begin{equation} \label{tfim}
    H(\lambda) =  J_{z}\sum_{i=1}^{N-1} \sigma_{i}^{z}\sigma_{i+1}^{z} + (1 + \frac{\lambda(t)}{2})h_x \sum_{i=1}^{N} \sigma_{i}^{x},
\end{equation}
where $N$ is the number of spins in the system, $J_z$ is the strength of the exchange interaction between pairs of nearest neighbor spins, $h_x$ is the strength of the transverse magnetic field, and $\sigma_i^{\alpha}$ is the $\alpha$-Pauli operator acting on spin $i$. The system starts in thermal equilibrium at an inverse temperature $\beta$ with an initial Hamiltonian $H_i = H(\lambda=0) = J_{z}\sum_{i} \sigma_{i}^{z}\sigma_{i+1}^{z} + h_x\sum_{i} \sigma_{i}^{x}$.  The parameter $\lambda$ is then linearly increased from $0$ to $1$ over a total time $tau$, resulting in a system with a final Hamiltonian $H_f = H(\lambda=1) = J_{z}\sum_{i} \sigma_{i}^{z}\sigma_{i+1}^{z} + 1.5h_x \sum_{i}\sigma_{i}^{x}$.  We set $J_z = 1$, $h_x = 1$, $\tau=10$, and set the number of trajectories $M = 100$ for the 2-qubit system and $M = 300$ for the 3-qubit system.  

Figure \ref{fig:hardware_results} shows the approximate free energy differences at various inverse temperatures $\beta$ computed using our algorithm on the IBM quantum processing unit (QPU) ``ibmq$\_$toronto" for a 2-qubit system (a) and a 3-qubit system (b).  The black solid lines show the analytically computed free energy differences, which are possible to compute due to the small size of our systems.  The blue dashed lines show raw results from the QPU.  The quantum circuits for the 3-qubit simulations are significantly larger than those for the 2-qubit simulations, and thus accumulate more error due to hardware noise.  This explains why the QPU results for the 2-qubit system are significantly closer to the ground truth than those for the 3-qubit system.  To ameliorate this systematic noise, we implement two post-processing error mitigation techniques.  The first is known as zero-noise extrapolation (ZNE) \cite{temme2017error, li2017efficient}, which combats noise arising from two-qubit entangling gates, which are currently one of the largest sources of error on NISQ devices.  We pair ZNE with a second error mitigation technique to combat readout error.  Readout error is related to error in the measurement operation. See Section VI of the SI for more details on error mitigation.  The QPU results after this post-processing are shown in the red dotted lines.  The error mitigated results are in excellent agreement with the analytic results for both size systems, demonstrating the ability of the two error mitigation techniques to combat major contributions to noise on the quantum computer.  

While error mitigation techniques can combat the noise present on near-term quantum devices, another source of error stems from the using the QITE algorithm to approximate the imaginary time evolution required to generate the METTS.  The size of this error depends on the step-size $\Delta \beta$ used to construct the QITE circuits for thermal state preparation at inverse temperature $\beta$.  This error can systematically be made smaller by decreasing $\Delta \beta$ at the expense of complexity in building the quantum circuit.  In general, Trotter error will be another source of error, which arises from the most commonly used approach to generate the real-time evolution operator $U(t)$ used to evolve the system as $\lambda$ is varied from 0 to 1.  However, we were able to make this negligible using techniques developed in \cite{bassman2021constantdepth}, which apply to real-time evolution of TFIMs.  See Section VII of the SI for more details.

\textit{Conclusion.--}We have introduced an algorithm for approximating free energy differences of quantum systems on quantum computers using fluctuation relations. We demonstrated our algorithm on IBM's quantum processor for the TFIM, resulting in free energy differences in excellent agreement with the ground truth after applying two simple error mitigation techniques.  The most resource intensive component of our algorithm is the generation of the initial thermal states, due to its dependence on the QITE algorithm.  We emphasize that other thermal state preparation algorithms~\cite{zhu2020generation} could potentially be substituted here in the future without changing the nature of our algorithm.  We anticipate this and related~\cite{francis2020many} algorithms to become increasingly important as a means to explore the thermodynamics of quantum systems as quantum hardware becomes more powerful.  

\textit{Acknowledgments.--}L.B., K.K., and W.A.d.J. were supported by the Office of Science, Office of Advanced Scientific Computing Research Quantum Algorithms Team and Accelerated Research for Quantum Computing Programs of the U.S. Department of Energy under Contract No. DE-AC02-05CH11231. D.L. was funded in part by the NSF Mathematical Sciences Graduate Internship Program. N.M.T. is grateful for support from NASA Ames Research Center and support from the AFRL Information Directorate under Grant No.~F4HBKC4162G001. This research used
resources of the Oak Ridge Leadership Computing Facility, which is a
DOE Office of Science User Facility supported under Contract
No.~DE-AC05-00OR22725. We are grateful to Yigit Subasi, Alexander Kemper, and Miles Stoudenmire for insightful discussions.

\newpage
\bibliographystyle{iopart-num}
\bibliography{main.bib}
\end{document}